\documentclass[12pt,onecolumn]{IEEEtran}

\usepackage{amssymb}
\usepackage{graphicx}
\usepackage{amsmath}
\usepackage[colorlinks,linkcolor=blue,citecolor=red,linktocpage=true]{hyperref}

\usepackage[pagewise]{lineno}

\begin {document}


\def\bbbf{{\rm I\!F}}

\def\bbbz{{\mathchoice {\hbox{$\sf\textstyle Z\kern-0.4em Z$}}
{\hbox{$\sf\textstyle Z\kern-0.4em Z$}}
{\hbox{$\sf\scriptstyle Z\kern-0.3em Z$}}
{\hbox{$\sf\scriptscriptstyle Z\kern-0.2em Z$}}}}

\newtheorem{definition}{Definition}
\newtheorem{proposition}{Proposition}
\newtheorem{theorem}{Theorem}

\newtheorem{remark}{Remark}
\newtheorem{example}{Example}

\newcommand{\Tr}{\textrm{\rm Tr}}
\newcommand{\Ord}{\textrm{\rm ord}}
\newcommand{\Sym}{\textrm{\rm SSymb}}
\newcommand{\Occ}{\#\textrm{\rm Occr}}


\title{Optimal three-weight cyclic codes whose duals are also optimal}

\author{
Gerardo Vega\thanks{G. Vega is with the Direcci\'on General de C\'omputo y de Tecnolog\'{\i}as de Informaci\'on y Comunicaci\'on, Uni\-ver\-si\-dad Nacional Aut\'onoma de 
M\'exico, 04510 M\'exico D.F., MEXICO (e-mail: gerardov@unam.mx).} and F\'elix Hern\'andez\thanks{F. Hern\'andez is a PhD student at the Posgrado en Ciencia e Ingenier\'{\i}a de la Computaci\'on, Universidad Nacional Aut\'onoma de M\'exico, 20059 M\'exico, D.F., MEXICO (e-mail: felixhdz@ciencias.unam.mx). Manuscript partially supported by CONACyT, M\'exico.}
}
\maketitle


\begin{abstract} 
A class of optimal three-weight cyclic codes of dimension 3 over any finite field was presented by Vega [Finite Fields Appl., 42 (2016) 23-38]. Shortly thereafter, Heng and Yue [IEEE Trans. Inf. Theory, 62(8) (2016) 4501-4513] generalized this result by presenting several classes of cyclic codes with either optimal three weights or a few weights. Here we present a new class of optimal three-weight cyclic codes of length $q+1$ and dimension 3 over any finite field $\bbbf_{q}$, and show that the nonzero weights are $q-1$, $q$, and $q+1$. We then study the dual codes in this new class, and show that they are also optimal cyclic codes of length $q+1$, dimension $q-2$, and minimum Hamming distance $4$. Lastly, as an application of the Krawtchouck polynomials, we obtain the weight distribution of the dual codes.
\end{abstract}

\noindent
{\it Keywords:} 
Optimal three-weight cyclic codes, optimal dual cyclic codes, Griesmer lower bound, weight distribution of a code.

\section{Introduction}\label{secuno}
The problem of obtaining the weight distribution of a given code is important because it plays a significant role in determining the capabilities of error detection and correction of such a code. For cyclic codes this problem is even more important because this kind of codes possess a rich algebraic structure (they are ideals in the principal ideal ring $\bbbf_{q}[x]/(x^n-1)$, where $n$ is the length of the cyclic codes). On the other hand, it is known that cyclic codes with few weights have a great practical importance in cryptography and coding theory since they are useful in the design of secret sharing schemes and association schemes (see \cite{Anderson,Calderbank}). A characterization of a class of optimal three-weight cyclic codes of dimension 3, over any finite field $\bbbf_{q}$, was presented in \cite{Vega1}, and shortly thereafter, several classes of cyclic codes with either optimal three weights or a few weights were given in \cite{Heng}, showing that one of these classes can be constructed as a generalization of the sufficient numerical conditions of the characterization given in \cite{Vega1}. 

In this paper we use a particular kind of one-weight and semiprimitive two-weight irreducible cyclic codes of dimension $2$ to construct a new class of optimal three-weight cyclic codes of length $q+1$ and dimension 3, over any finite field $\bbbf_{q}$, whose nonzero weights correspond to the three largest possible weights, that is $q-1$, $q$, and $q+1$. The codes in this class are optimal in the sense that their lengths reach the Griesmer lower bound for linear codes. Furthermore, without the need of any exponential sum, we explicitly determine the weight distribution for the cyclic codes in this class. With the knowledge of this weight distribution, we then study the corresponding dual codes showing that, except for a single case, all of them have minimum Hamming distance $4$. In consequence, since the length, dimension and minimum Hamming distance of the dual codes are $q+1$, $q-2$, and 4, respectively, we conclude that these codes are also optimal. As an application of the Krawtchouck polynomials, we obtain the weight distribution of the dual codes, in which it is clear that all of them are $(q-2)$-weight cyclic codes for $q\geq 5$. Therefore, in addition to our new class of optimal cyclic codes, this last result gives access to an infinite family of optimal $(q-2)$-weight cyclic codes of length $q+1$, and dimension $q-2$, whose minimum Hamming distance is $4$.

This work is organized as follows: In Section \ref{secdos} we establish some notation, recall some definitions and already known results related to the weight distributions of one-weight and semiprimitive two-weight irreducible cyclic codes of dimension $2$. In addition, we also recall the Griesmer lower bound for linear codes, the five first identities of Pless, and the definition for the Krawtchouck polynomials. In Section \ref{sectres} some preliminary results are presented. Particularly, a description of the nature of the codewords in either a one-weight or a semiprimitive two-weight irreducible cyclic code of dimension $2$. This description is then used in Section \ref{seccuatro} to present a new class of optimal three-weight cyclic codes of dimension 3 over any finite field, showing that their dual codes are also optimal. Examples of optimal three-weight cyclic codes belonging to this new class, along with their corresponding dual codes, are presented at the end of this section. Finally, Section \ref{conclusiones} is devoted to conclusions.

\section{Definitions, notation and some known results}\label{secdos}

Unless otherwise specified, throughout this work we are going to use the following:

\medskip
\noindent
{\bf Notation.} By using $q$ and $k$, we will denote positive integers such that $q$ is the power of a prime number. For integers $v$ and $w$, with $\gcd(v,w)=1$, $\mbox{ord}_v(w)$ will denote the {\em multiplicative order} of $w$ modulo $v$. ``$\mbox{Tr}_{\bbbf_{q^k}/\bbbf_q}$" will denote the trace mapping from $\bbbf_{q^k}$ to $\bbbf_q$. By using $\gamma$, we will denote a fixed primitive element of $\bbbf_{q^k}$, and for any integer $a$, the polynomial $h_a(x) \in \bbbf_{q}[x]$ will denote the {\em minimal polynomial} of $\gamma^{-a}$ (see, for example, \cite[p. 99]{MacWilliams}). In addition, ${\cal C}_{(n,q^k)}$ will denote the irreducible cyclic code of length $n=\frac{q^k-1}{\gcd(q^k-1,a)}$, whose parity-check polynomial is $h_a(x)$. Note that ${\cal C}_{(n,q^k)}$ is an $[n,k']$ linear code, where its dimension $k'=\deg(h_a(x))=\mbox{ord}_n(q)$ is a divisor of $k$. Note also that ${\cal C}_{(n,q^k)}$ is not several repetitions of an irreducible cyclic code of smaller block length.

Let ${\cal C}$ be a linear code of length $n$ over $\bbbf_q$. Then, the dual code, ${\cal C}^{\perp}$, of ${\cal C}$ is the linear code defined by 

$${\cal C}^{\perp}:=\{v \in \bbbf_{q}^n \:|\: \langle v,c \rangle=0, \mbox{ for all } c \in {\cal C}\}\;,$$

\noindent
where $\langle v,c \rangle$ is a scalar product in the vector space $\bbbf_{q}^n$. It is known that if ${\cal C}$ is an $[n,k]$ linear code, then its dual code, ${\cal C}^{\perp}$, is an $[n,n-k]$ linear code. For $0\leq i \leq n$, $A_i$ and $A_i^{\perp}$ will denote the number of codewords of weight $i$ in ${\cal C}$ and in ${\cal C}^{\perp}$, respectively. 

An $M$-\textit{weight} code is a code such that the cardinality of the set of nonzero weights is $M$. That is, $M=|\{i:A_i\neq0,i=1,2,3,\ldots,n\}|$. 

An alternative definition for an irreducible cyclic code is as follows:

\begin{definition}\label{defICC} 
\cite[Definition 2.2]{Schmidt} Let $n$ be a positive divisor of $q^k-1$. Then, an irreducible cyclic code, ${\cal C}_{(n,q^k)}$, of length $n$ and dimension $\Ord_n(q)$, over $\bbbf_{q}$, is the set 

\[{\cal C}_{(n,q^k)}:=\{ c(\beta) \: | \: \beta \in \bbbf_{q^k} \} \; ,\]

\noindent
where 

\[c(\beta):=(\Tr_{\bbbf_{q^k}/\bbbf_q}(\beta \gamma^{\frac{q^k-1}{n}i}))_{i=0}^{n-1} \;.\]
\end{definition} 

\begin{remark}\label{rmcero} 
Note that, thanks to Delsarte's Theorem (see, for example, \cite{Delsarte}), the parity-check polynomial of the irreducible cyclic code under the previous definition is $h_{-\frac{q^k-1}{n}}(x)$. 
\end{remark}

Now, for the particular case of $k=2$, let $n_1$ and $n_2$ be divisors of $q^2-1$ and assume that $n_1 | (q-1)$ and $n_2 \nmid (q-1)$. Note that under these circumstances, $\gamma^{q+1}$ is a primitive element of $\bbbf_{q}$, $\Ord_{n_1}(q)=1$, and $\Ord_{n_2}(q)=2$. With this in mind, we present the following:

\begin{definition}\label{defRCC} 
Let $n_1$ and $n_2$ be as before, and let $n=(q^2-1)/\gcd(\frac{q^2-1}{n_1},\frac{q^2-1}{n_2})$. Then, a reducible cyclic code of length $n$ and dimension $3$, ${\cal C}_{(n_1,n_2,q^2)}$, over $\bbbf_{q}$, is given by the set:

\[{\cal C}_{(n_1,n_2,q^2)}:=\{ c(\alpha,\beta) \: | \: \alpha \in \bbbf_q, \: \beta \in \bbbf_{q^2} \} \; ,\]

\noindent
where 

\[c(\alpha,\beta):=(\alpha (\gamma^{q+1})^{\frac{q-1}{n_1}i}+\Tr_{\bbbf_{q^2}/\bbbf_q}(\beta \gamma^{\frac{q^2-1}{n_2}i}))_{i=0}^{n-1} \;.\]
\end{definition} 

When constructing a code, from an economical point of view, it is desirable to obtain an $[n,k,d]$ code ${\cal C}$ over $\bbbf_q$ whose length $n$ is minimal for given values of $k$, $d$ and $q$. A lower bound for the length $n$ in terms of these values is as follows. Let $n_q(k,d)$ be the minimum length $n$ for which an $[n,k,d]$ linear code, over $\bbbf_{q}$, exists. If the values of $q$, $k$ and $d$ are given, then a well-known lower bound (see \cite{Griesmer} and \cite{Solomon}) for $n_q(k,d)$ is:

\begin{theorem}\label{teoGriesmer}
(Griesmer bound) With the previous notation,

\[n_q(k,d) \geq \sum_{i=0}^{k-1} \left \lceil \frac{d}{q^i} \right \rceil \; ,\]

\noindent
where $\left \lceil x \right \rceil$ denotes the smallest integer greater than or equal to $x$.
\end{theorem} 

Let ${\cal C}$ be a linear code of length $n$ and dimension $k$ over $\bbbf_{q}$. With our notation, fix $m=n(q-1)$, and suppose that the weight of the dual of ${\cal C}$ is at least 2 (that is, $A_1^{\perp}=0$). Then, the five first identities of Pless (see either \cite[pp. 259-260]{Huffman} or \cite{Pless} for the general result), for ${\cal C}$, are:

\begin{eqnarray}\label{eqPless}
\sum_{i=1}^n A_i &=& q^k-1 \;, \nonumber \\
\sum_{i=1}^n i A_i &=& q^{k-1}m \;, \nonumber \\
\sum_{i=1}^n i^2 A_i &=& q^{k-2}[m(m+1)+2A_2^{\perp}] \;, \nonumber \\
\sum_{i=1}^n i^3 A_i &=& q^{k-3}[m(m(m+3)-q+2)+6(m-q+2)A_2^{\perp}-6A_3^{\perp}] \;, \nonumber \\
\sum_{i=1}^n i^4 A_i &=& q^{k-4}[m(m(m(m+6)-4q+11)+q^2-6(q-1)) \nonumber \\
&& +(12m(m-2q+5)+14q^2-72(q-1))A_2^{\perp} \nonumber \\
&& -(24m-36(q-2))A_3^{\perp}+24A_4^{\perp}] \;.
\end{eqnarray}

When the weight distribution of a linear code is known, it is possible to obtain the weight distribution of its corresponding dual code using the Krawtchouck polynomials. Through the following, we recall such an important mathematical tool (see, for example, \cite[pp. 256]{Huffman}).

\begin{definition}\label{defKP} 
Let $q$, $n$, and $j$ be positive integers such that $q$ is a prime power. Then, we define the {\em Krawtchouck polynomial}, $K_j^{n,q}(x)$, of degree $j$ to be

\[K_j^{n,q}(x):=\sum_{l=0}^j (-1)^l (q-1)^{j-l}\left(\begin{array}{c} x \\ l \\ \end{array} \right) \left(\begin{array}{c} n-x \\ j-l \\ \end{array} \right)\; .\]
\end{definition} 

Let $A_i$ and $A_i^{\perp}$ be as before. Then, for a given linear code ${\cal C}$ of length $n$, the weight distribution of its dual code can be obtained by means of

\begin{equation}\label{eqdual1}
A_j^{\perp}=\frac{1}{|{\cal C}|}\sum_{i=0}^n A_i K_j^{n,q}(i)\;,\;\;\; 0\leq j\leq n\;.
\end{equation}

It was shown in \cite{Vega2} that all irreducible cyclic codes of dimension $1$ or $2$ are either one-weight or semiprimitive two-weight irreducible cyclic codes. Since such a result is important for this work, we recall it by means of the following:

\begin{theorem}\label{teoCCI}
Let $n$ be a positive divisor of $q^2-1$. Fix $u=\gcd(q+1,\frac{q^2-1}{n})$. In accordance with Definition \ref{defICC}, let ${\cal C}_{(n,q^2)}$ be the irreducible cyclic code of length $n$ over $\bbbf_{q}$. Thus, we have the following assertions.

\begin{enumerate}
\item[\rm (a)] If $1 \leq u < q+1$, then ${\cal C}_{(n,q^2)}$ is either a one-weight or a semiprimitive two-weight irreducible cyclic code of dimension $2$, whose weight enumerator polynomial is

\[ 1+\frac{(q^2-1)}{u}z^{\frac{n(q+1-u)}{q+1}}+\frac{(q^2-1)(u-1)}{u}z^{n} \; . \] 

\item[\rm (b)] If $u=q+1$, then ${\cal C}_{(n,q^2)}$ is an $[n,1]$ one-weight code, whose nonzero weight is $n$ (which is equivalent to a repetition code of length $n$).
\end{enumerate}
\end{theorem}

\section{Some preliminary results}\label{sectres}

In this section, and for the rest of the paper, we will assume that $\gamma$ is a fixed primitive element of $\bbbf_{q^2}$.

The following two definitions will be important in order to achieve our goals.

\begin{definition}\label{defuno}
Let $V=(v_0,v_1,\cdots,v_{n-1})$ be a vector of length $n$ over $\bbbf_{q}$. We define {\em the set of symbols of $V$}, $\Sym(V)$, as the subset of symbols of $\bbbf_{q}$ appearing as entries in the vector $V$. That is:

$$\Sym(V=(v_0,v_1,\cdots,v_{n-1})):=\{ s \in \bbbf_{q} \:|\: s=v_i, \mbox{ for some } 0\leq i < n\}\;.$$
\end{definition}

\begin{definition}\label{defdos}
Let $s \in \bbbf_{q}$, and let $V=(v_0,v_1,\cdots,v_{n-1})$ be a vector of length $n$ over $\bbbf_{q}$. We define {\em the number of occurrences of the symbol $s$ in $V$}, $\Occ(s,V)$, as the number of times that $s$ appears as an entry in the vector $V$. That is:

$$\Occ(s,V=(v_0,v_1,\cdots,v_{n-1})):=|\{ i \:|\: s=v_i, \:0\leq i < n\}|\;.$$
\end{definition}

\begin{remark}\label{rmuno} 
Note that $0 \leq \Occ(s,V) \leq n$. In fact, clearly, $\Occ(s,V)>0$ iff $s \in \Sym(V)$.
\end{remark}

The following is an easy result whose proof is given here in order to make this paper as self-contained as possible.

\begin{proposition}\label{ProTrz}
Let $l$ be an integer. If $q$ is odd, then $\Tr_{\bbbf_{q^2}/\bbbf_q}(\gamma^{\frac{q+1}{2}+l(q+1)})=0$. In addition, if $q$ is even, then $\Tr_{\bbbf_{q^2}/\bbbf_q}(\gamma^{l(q+1)})=0$.
\end{proposition}

\begin{proof} 
First note that $\gamma^{l(q+1)} \in \bbbf_{q}^*$. Thus, if $q$ is odd, then 

$$\Tr_{\bbbf_{q^2}/\bbbf_q}(\gamma^{\frac{q+1}{2}+l(q+1)})=\gamma^{l(q+1)}\Tr_{\bbbf_{q^2}/\bbbf_q}(\gamma^{\frac{q+1}{2}})\;,$$

\noindent
but,

$$\Tr_{\bbbf_{q^2}/\bbbf_q}(\gamma^{\frac{q+1}{2}})=(\gamma^{\frac{q+1}{2}}+\gamma^{\frac{q(q+1)}{2}})=(\gamma^{\frac{q+1}{2}}+\gamma^{\frac{q^2-1}{2}}\gamma^{\frac{q+1}{2}})=(\gamma^{\frac{q+1}{2}}-\gamma^{\frac{q+1}{2}})=0\;.$$

On the other hand, if $q$ is even, then 

$$\Tr_{\bbbf_{q^2}/\bbbf_q}(\gamma^{l(q+1)})=\gamma^{l(q+1)}+\gamma^{l(q+1)}=0\;.$$
\end{proof}

\begin{remark}\label{rmCR} 
Note that in the case when $n_1=1$ and $n_2=q+1$, in Definition \ref{defRCC}, the codewords $c(\alpha,\beta)$, in the reducible cyclic code ${\cal C}_{(1,q+1,q^2)}$, are of the form:

\[c(\alpha,\beta)=(\alpha +\Tr_{\bbbf_{q^2}/\bbbf_q}(\beta \gamma^{(q-1)i}))_{i=0}^{q} \;, \mbox{ with } \alpha \in \bbbf_q,\; \beta \in \bbbf_{q^2} \;. \]

\noindent
In fact, in this case, it is important to keep in mind that ${\cal C}_{(1,q+1,q^2)}$ is a reducible cyclic code of length $q+1$ and dimension $3$. In addition, observe that the reducible cyclic code ${\cal C}_{(1,q+1,q^2)}$ is the sum, as vector spaces, of the irreducible cyclic codes ${\cal C}_{(1,q^2)}$ and ${\cal C}_{(q+1,q^2)}$. Lastly, note that, owing to Part {\rm (b)} of Theorem \ref{teoCCI}, ${\cal C}_{(1,q^2)}$ is a $[1,1]$ one-weight irreducible cyclic code, while, owing to Part {\rm (a)} of the same theorem, ${\cal C}_{(q+1,q^2)}$ is either a one-weight or a semiprimitive two-weight irreducible cyclic code of length $q+1$ and dimension $2$.  
\end{remark}

The following result is key to identify the number of occurrences of a given symbol $s$ in a codeword $c(\beta)$ that belongs to an irreducible cyclic code of the form ${\cal C}_{(q+1,q^2)}$ (see Definition \ref{defICC}).

\begin{proposition}\label{pruno}
Let $b$, $j$ and $t$ be integers such that $0 < t < (q+1)$. Then, $\Tr_{\bbbf_{q^2}/\bbbf_q}(\gamma^{b}\gamma^{(q-1)j})=\Tr_{\bbbf_{q^2}/\bbbf_q}(\gamma^{b}\gamma^{(q-1)(j+t)})$ iff $(q+1) | (2j+t-b)$.
\end{proposition}

\begin{proof} 
$\Tr_{\bbbf_{q^2}/\bbbf_q}(\gamma^{b}\gamma^{(q-1)j})=\Tr_{\bbbf_{q^2}/\bbbf_q}(\gamma^{b}\gamma^{(q-1)(j+t)})$ iff

$$\gamma^{(q-1)j+b}+\gamma^{q((q-1)j+b)}=\gamma^{(q-1)(j+t)+b}+\gamma^{q((q-1)(j+t)+b)}\;.$$

\noindent
But $\gamma^{q(q-1)x}=\gamma^{-(q-1)x}$, for all integer $x$. Therefore,

\vspace{-0.08in}
\begin{eqnarray}
\gamma^{(q-1)j+b}+\gamma^{-(q-1)j+qb}&=&\gamma^{(q-1)(j+t)+b}+\gamma^{-(q-1)(j+t)+qb}\;, \nonumber \\
\gamma^{(q-1)j+b}-\gamma^{(q-1)(j+t)+b}&=&\gamma^{-(q-1)(j+t)+qb}-\gamma^{-(q-1)j+qb}\;, \nonumber \\
\gamma^{(q-1)j}-\gamma^{(q-1)(j+t)}&=&\gamma^{-(q-1)(j+t)+(q-1)b}-\gamma^{-(q-1)j+(q-1)b}\;, \nonumber \\
\gamma^{(q-1)j}(1-\gamma^{(q-1)t})&=&\gamma^{-(q-1)(j+t)+(q-1)b}(1-\gamma^{(q-1)t})\;. \nonumber
\end{eqnarray}

\noindent
Since $0 < t < (q+1)$, $(1-\gamma^{(q-1)t})\neq 0$. Therefore,

\vspace{-0.08in}
\begin{eqnarray}
\gamma^{(q-1)j}&=&\gamma^{-(q-1)(j+t)+(q-1)b}\;, \nonumber \\
\gamma^{(q-1)2j}&=&\gamma^{(q-1)(b-t)}\;. \nonumber
\end{eqnarray}

\noindent
Now, note that the previous equality will hold iff $2j \equiv b-t \pmod{q+1}$ iff $(q+1)|(2j+t-b)$.
\end{proof}

In the light of Remark \ref{rmCR}, we can now describe the nature of a codeword $c(\beta)$ that belongs to an irreducible cyclic code of the form ${\cal C}_{(q+1,q^2)}$.

\begin{proposition}\label{prdos}
Let $\beta \in \bbbf_{q^2}^*$, and let $b$ be the index of $\beta$ (that is, $\beta=\gamma^b$). Let $j$ be an integer. Then, for the codeword $c(\beta)=\{(\Tr_{\bbbf_{q^2}/\bbbf_q}(\gamma^b \gamma^{(q-1)i}))_{i=0}^{q}\} \in {\cal C}_{(q+1,q^2)}$, we have

\begin{enumerate}
\item[\rm (a)] If $\gamma^b \gamma^{(q-1)j} \in \bbbf_{q}^*$, then $\Occ(\Tr_{\bbbf_{q^2}/\bbbf_q}(\gamma^b \gamma^{(q-1)j}),c(\beta))=1$.

\item[\rm (b)] If $\gamma^b \gamma^{(q-1)j} \notin \bbbf_{q}^*$, then $\Occ(\Tr_{\bbbf_{q^2}/\bbbf_q}(\gamma^b \gamma^{(q-1)j}),c(\beta))=2$. 

\item[\rm (c)] For all $s \in \bbbf_q$, $\Occ(s,c(\beta)) \leq 2$.

\item[\rm (d)] Let $s \in \bbbf_q^*$. Then, 

$$|\{ c(\beta) \in {\cal C}_{(q+1,q^2)} \:|\: \Occ(s,c(\beta))=1\}|=\left\{ \begin{array}{cl}
		\!q+\!1 & \!\mbox{if $q$ is odd,} \\
\\
		\!0 & \!\mbox{otherwise. }
			\end{array}
\right . $$

\item[\rm (e)] Let $s \in \bbbf_q$, such that $\Occ(s,c(\beta))=1$. Then, $q$ is odd iff $s\neq 0$.

\item[\rm (f)] Let $s \in \bbbf_q$, such that $\Occ(s,c(\beta))=2$. If $q$ is even, then $s\neq 0$.

\end{enumerate}
\end{proposition}

\begin{proof} 
Parts {\rm (a)} and {\rm (b)}: First, note that because $(q+1)|(2j+(q-1)j)$, we have $\gamma^b \gamma^{(q-1)j} \in \bbbf_{q}^*=\langle \gamma^{q+1} \rangle$ iff $(q+1)|((q-1)j+b)$ iff $(q+1)|(2j-b)$. Thus, if $(q+1) | (2j-b)$, then there is no integer $0 < t < (q+1)$ such that $(q+1) | (2j+t-b)$, which in turn implies, due to Proposition \ref{pruno}, that there is no integer $0 < t < (q+1)$ such that $\Tr_{\bbbf_{q^2}/\bbbf_q}(\gamma^{b}\gamma^{(q-1)j})=\Tr_{\bbbf_{q^2}/\bbbf_q}(\gamma^{b}\gamma^{(q-1)(j+t)})$. Therefore, $\Occ(\Tr_{\bbbf_{q^2}/\bbbf_q}(\gamma^{b}\gamma^{(q-1)j}),c(\beta))=1$. On the contrary, if $(q+1) \nmid (2j-b)$, then there exists a unique integer $0 < t < (q+1)$ such that $(q+1) | (2j+t-b)$, and for such an integer $t$ we have $\Tr_{\bbbf_{q^2}/\bbbf_q}(\gamma^{b}\gamma^{(q-1)j})=\Tr_{\bbbf_{q^2}/\bbbf_q}(\gamma^{b}\gamma^{(q-1)(j+t)})$. Therefore, in this case, $\Occ(\Tr_{\bbbf_{q^2}/\bbbf_q}(\gamma^{b}\gamma^{(q-1)j}),c(\beta))=2$.

Part {\rm (c)}: If $s \notin \Sym(c(\beta))$, then, clearly, $\Occ(s,c(\beta))=0$. On the other hand, if $s \in \Sym(c(\beta))$, then Parts {\rm (a)} and {\rm (b)} show that $\Occ(s,c(\beta))$ is either $1$ or $2$.

Part {\rm (d)}: Suppose that $q$ is odd. Let $0\leq r < q-1$ be the unique integer such that $2^{-1}s=\gamma^{(q+1)r} \in \bbbf_q^*$, and observe that if $\gamma^{b'}\gamma^{(q-1)j}=\gamma^{(q+1)r}$, for integers $j$ and $b'$, then $\Tr_{\bbbf_{q^2}/\bbbf_q}(\gamma^{b'} \gamma^{(q-1)j})=2^{-1}s\Tr_{\bbbf_{q^2}/\bbbf_q}(1)=s$. That is, since $\gamma^{b'} \gamma^{(q-1)j} \in \bbbf_{q}^*$, $\Occ(s,c(\gamma^{b'}))=1$. Then, the result follows from the fact that for each $0\leq j \leq q$ there exists a unique integer $0\leq b' < q^2-1$ such that $\gamma^{b'}\gamma^{(q-1)j}=\gamma^{(q+1)r}$.

On the contrary, if $q$ is even, then Parts {\rm (a)} and {\rm (b)}, and Proposition \ref{ProTrz}, show that if $s\in \Sym(c(\beta))$, then $\Occ(s,c(\beta))=1$ iff $s=0$. Thus, the result follows from the fact that it is impossible to find a codeword $c(\beta)$ such that $s \in \bbbf_q^*$, and $\Occ(s,c(\beta))=1$.

Parts {\rm (e)} and {\rm (f)}: Note that $\gamma^b \gamma^{(q-1)j} \in \bbbf_{q}^*$ iff $(q+1) | (b+(q-1)j)$. Thus, these results follow from Parts {\rm (a)} and {\rm (b)}, and Proposition \ref{ProTrz}.
\end{proof}
 
\begin{remark}\label{rmOcc} 
Note that the proof of the previous result ensures the existence of at least two codewords, $c(\beta)$ and $c(\beta')$ in ${\cal C}_{(q+1,q^2)}$, such that $\Occ(s,c(\beta))=1$ and $\Occ(s',c(\beta'))=2$, for some symbols $s$ and $s'$ such that $s \in \Sym(c(\beta))$ and $s' \in \Sym(c(\beta'))$.
\end{remark}

\begin{example}\label{ejeuno}
Let $\bbbf_7=\bbbz_7=\{0,1,\cdots,6\}$, and $\bbbf_{7^2}=\bbbf_7(\gamma),\;\gamma^2+6\gamma+3=0$. Then, for example, $c(\gamma^0)=(2,3,0,4,5,4,0,3)$ and $c(\gamma^1)=(1,1,2,5,6,6,5,2)$. Therefore, we have $\Occ(2,c(\gamma^0))=\Occ(5,c(\gamma^0))=1$, $\Occ(s,c(\gamma^0))=2$ for all $s \in \Sym(c(\gamma^0)) \setminus \{2,5\}$, and $\Occ(s,c(\gamma^1))=2$ for all $s \in \Sym(c(\gamma^1))$.
\end{example}

\begin{example}\label{ejedos}
Let $\bbbf_8=\{\hat{0},\hat{1},\cdots,\hat{7}\}$, where $\bbbf_8=\bbbf_2(\alpha),\;\alpha^3+\alpha+1$, and $\hat{0}:=0$, $\hat{1}:=1$, $\hat{2}:=\alpha$, $\hat{3}:=\alpha+1$, $\hat{4}:=\alpha^2$, $\hat{5}:=\alpha^2+1$, $\hat{6}:=\alpha^2+\alpha$, and $\hat{7}:=\alpha^2+\alpha+1$. By taking $\bbbf_{8^2}=\bbbf_8(\gamma),\;\gamma^2+\gamma+\hat{3}=0$, we have, for example, $c(\gamma^0)=(\hat{0},\hat{6},\hat{2},\hat{1},\hat{4},\hat{4},\hat{1},\hat{2},\hat{6})$ and $c(\gamma^1)=(\hat{1},\hat{1},\hat{7},\hat{5},\hat{4},\hat{0},\hat{4},\hat{5},\hat{7})$. Therefore, we have $\Occ(\hat{0},c(\gamma^0))=\Occ(\hat{0},c(\gamma^1))=1$, while $\Occ(s,c(\gamma^0))=\Occ(s',c(\gamma^1))=2$ for all $\hat{0} \neq s \in \Sym(c(\gamma^0))$, and $\hat{0} \neq s' \in \Sym(c(\gamma^1))$.
\end{example}

By means of the following two results we determine the nonzero weights in the reducible cyclic code ${\cal C}_{(1,q+1,q^2)}$, and also obtain the frequency of occurrence of one of these nonzero weights.

\begin{proposition}\label{prtres}
Let $\bar{1}=(1,1,\cdots,1)$ be the all-ones vector of length $q+1$. Then,

\[|\{ w_H(\alpha\bar{1} + c(\beta))=q \;|\; \alpha \in \bbbf_{q}^*, \beta \in \bbbf_{q^2}^*, c(\beta) \in {\cal C}_{(q+1,q^2)} \}|= \left\{ \begin{array}{cl}
		\!q^2\!-\!1 & \!\mbox{if $q$ is odd,} \\
\\
		\!0 & \!\mbox{otherwise, }
			\end{array}
\right .\]

\noindent
where $w_H(\cdot)$ stands for the usual Hamming weight function.
\end{proposition}

\begin{proof} 
Case 1: $q$ is odd. Let $s \in \bbbf_q$, such that $\Occ(s,c(\beta))=1$. Thus, owing to Part {\rm (e)} of Proposition \ref{prdos}, $s\neq 0$. By considering Part {\rm (d)} of such a proposition, the result now follows from the facts that $|\bbbf_{q}^*|=q-1$ and 

$$|\{ w_H((-s)\bar{1} + c(\beta))=q \;|\; \beta \in \bbbf_{q^2}^*, c(\beta) \in {\cal C}_{(q+1,q^2)} \}|=q+1\;.$$

Case 2: $q$ is even. In this case, Part {\rm (d)} of Proposition \ref{prdos} shows that 

$$|\{ w_H((-s)\bar{1} + c(\beta))=q \;|\; \beta \in \bbbf_{q^2}^*, c(\beta) \in {\cal C}_{(q+1,q^2)} \}|=0\;,$$

\noindent
and the result follows now directly.
\end{proof}

\begin{proposition}\label{prFrec}
With our notation, ${\cal C}_{(1,q+1,q^2)}$ is a three-weight cyclic code of length $q+1$ and dimension $3$, whose nonzero weights are $q-1$, $q$, and $q+1$. In addition, if $A_q$ denotes the number of codewords of weight $q$ in ${\cal C}_{(1,q+1,q^2)}$, then $A_q=q^2-1$.
\end{proposition}

\begin{proof} 
By Definition \ref{defRCC}, ${\cal C}_{(1,q+1,q^2)}$ is a reducible cyclic code of length $q+1$ and dimension $3$.

Case 1: $q$ is odd. In this case, $u=\gcd(q+1,\frac{q^2-1}{q+1})=\gcd(q+1,q-1)=2$. Therefore, owing to Part {\rm (a)} of Theorem \ref{teoCCI}, ${\cal C}_{(q+1,q^2)}$ is a semiprimitive two-weight irreducible cyclic code of length $q+1$ and dimension $2$, whose weight enumerator polynomial is $1+\frac{(q^2-1)}{2}z^{q-1}+\frac{(q^2-1)}{2}z^{q+1}$. That is, the two nonzero weights of ${\cal C}_{(q+1,q^2)}$ are $q-1$ and $q+1$. Since ${\cal C}_{(q+1,q^2)} \subsetneq {\cal C}_{(1,q+1,q^2)}$ (see Remark \ref{rmCR}), these two nonzero weights also belong to ${\cal C}_{(1,q+1,q^2)}$. Now, owing to Remark \ref{rmOcc} and Part {\rm (e)} of Proposition \ref{prdos}, there exist a codeword $c(\beta) \in {\cal C}_{(q+1,q^2)}$ such that $\Occ(s,c(\beta))=1$ for some symbol $s \neq 0$. Thus, note that $(-s)\bar{1}+c(\beta) \in {\cal C}_{(1,q+1,q^2)}$ and $w_H((-s)\bar{1}+c(\beta))=q$. On the other hand, by means of Part {\rm (c)} of Proposition \ref{prdos}, we can see that the only nonzero weights of ${\cal C}_{(1,q+1,q^2)}$ are $q-1$, $q$, and $q+1$. Lastly, Proposition \ref{prtres} shows now that $A_q=q^2-1$. 

Case 2: $q$ is even. In this case, $u=\gcd(q+1,q-1)=1$. Therefore, owing to Part {\rm (a)} of Theorem \ref{teoCCI}, ${\cal C}_{(q+1,q^2)}$ is a one-weight irreducible cyclic code of length $q+1$ and dimension $2$, whose nonzero weight is $q$. This means that the weight $q$ appears in ${\cal C}_{(q+1,q^2)}$, $q^2-1$ times. Now, since $\bar{1} \in {\cal C}_{(1,q+1,q^2)}$ (see Remark \ref{rmCR}), $q$ and $q+1$ are two nonzero weights in ${\cal C}_{(1,q+1,q^2)}$. But, Remark \ref{rmOcc} and Part {\rm (f)} of Proposition \ref{prdos}, show that there exist a codeword $c(\beta) \in {\cal C}_{(q+1,q^2)}$ such that $\Occ(s,c(\beta))=2$, for some symbol $s\neq 0$. Therefore, note that $(-s)\bar{1}+c(\beta) \in {\cal C}_{(1,q+1,q^2)}$ and $w_H((-s)\bar{1}+c(\beta))=q-1$. On the other hand, Part {\rm (c)} of Proposition \ref{prdos} shows that the only nonzero weights of ${\cal C}_{(1,q+1,q^2)}$ are $q-1$, $q$, and $q+1$. Lastly, owing to Proposition \ref{prtres}, $A_q=q^2-1$. 
\end{proof}

\section{Optimal three-weight cyclic codes of dimension 3 and their duals}\label{seccuatro}

We are now able to present our main results.

\begin{theorem}\label{mainR1}
With our notation, ${\cal C}_{(1,q+1,q^2)}$ is an optimal three-weight cyclic code of length $q+1$ and dimension $3$, whose nonzero weights are $q-1$, $q$, and $q+1$. In addition, the weight enumerator polynomial of ${\cal C}_{(1,q+1,q^2)}$ is

\[ 1+\frac{q(q^2-1)}{2}z^{q-1}+(q^2-1)z^{q}+\frac{q(q-1)^2}{2}z^{q+1} \; . \] 
\end{theorem}

\begin{proof} 
By Proposition \ref{prFrec}, ${\cal C}_{(1,q+1,q^2)}$ is a three-weight cyclic code of length $q+1$ and dimension $3$, whose nonzero weights are $q-1$, $q$, and $q+1$.

The dimension, $k$, and minimum distance, $d$, of ${\cal C}_{(1,q+1,q^2)}$ are $3$ and $q-1$, respectively. Therefore, a direct application of Theorem \ref{teoGriesmer} shows that 

\[\sum_{i=0}^{2} \left \lceil \frac{q-1}{q^i} \right \rceil = q-1+\sum_{i=1}^{2}(1)=q-1 + (2)=q+1\; .\]

\noindent
Consequently, ${\cal C}_{(1,q+1,q^2)}$ is an optimal linear code in the sense that its length, $q+1$, reaches the lower bound in such a theorem.

It is well known that the minimum Hamming distance of the dual of any nonzero cyclic code is greater than 1 (see, for example, \cite[Section V]{Wolfmann}). On the other hand, Proposition \ref{prFrec} tells us that $A_q=q^2-1$. Thus, by using the first two identities in (\ref{eqPless}), we obtain the following two linear equations:

\begin{eqnarray}
A_{q-1}+(q^2-1)+A_{q+1}&=&q^3-1 \;, \nonumber \\
A_{q-1}(q-1)+(q^2-1)q+A_{q+1}(q+1)&=&(q+1)(q-1)q^2 \;. \nonumber
\end{eqnarray}

\noindent
The solution of such equations shows that $A_{q-1}=\frac{q(q^2-1)}{2}$ and $A_{q+1}=\frac{q(q-1)^2}{2}$. Therefore, the weight enumerator polynomial for ${\cal C}_{(1,q+1,q^2)}$ is the required one.
\end{proof}

\begin{theorem}\label{mainR2}
Denote by ${\cal C}_{(1,q+1,q^2)}^{\perp}$ the dual code of ${\cal C}_{(1,q+1,q^2)}$. Suppose that $\bbbf_{q}$ is not the binary field (that is, $q\neq 2$). Then, ${\cal C}_{(1,q+1,q^2)}^{\perp}$ is an optimal cyclic code of length $q+1$, dimension $q-2$, and minimum Hamming distance $4$. In fact, 

$$A_4^{\perp}=\frac{q(q^2-1)(q-1)(q-2)}{24}\;.$$ 

\noindent
In addition, if $q\geq 5$, then ${\cal C}_{(1,q+1,q^2)}^{\perp}$ is a $(q-2)$-weight cyclic code whose weight distribution is given by

\begin{equation}\label{eqdual2}
A_j^{\perp}=\frac{(q^2-1)\left(\begin{array}{c} q-1 \\ j-2 \\ \end{array} \right)}{2j(j-1)q^2}\left[2(q-1)^{j-1}+(-1)^j((j-1)q((j-2)q-2)+2)\right]\;,\end{equation}

\noindent
for $4\leq j \leq q+1$.
\end{theorem}

\begin{proof} 
Clearly, ${\cal C}_{(1,q+1,q^2)}^{\perp}$ is a cyclic code of length $q+1$ and dimension $(q+1)-3=q-2$. On the other hand, if $q\neq 2$, then the dual code ${\cal C}_{(1,q+1,q^2)}^{\perp}$ is not the null code $\{0\}$. Therefore, when $q\neq 2$, a direct application of the last three identities in (\ref{eqPless}), shows that $A_2^{\perp}=A_3^{\perp}=0$, and that the value of $A_4^{\perp}$ is the announced one. On the other hand, the Griesmer lower bound shows that

\[\sum_{i=0}^{q-3} \left \lceil \frac{4}{q^i} \right \rceil = 4+\sum_{i=1}^{q-3}(1)=4 + (q-3)=q+1\; .\]

\noindent
Therefore, the dual code ${\cal C}_{(1,q+1,q^2)}^{\perp}$, like ${\cal C}_{(1,q+1,q^2)}$, is an optimal cyclic code of length $q+1$. Now, through Definition \ref{defKP}, it is not difficult to see that

\begin{eqnarray}
K_j^{q+1,q}(0) &=& \frac{q\left(\begin{array}{c} q-1 \\ j-2 \\ \end{array} \right)}{j(j-1)} (q^2-1)(q-1)^{j-1}\;, \nonumber \\
K_j^{q+1,q}(q-1) &=& \frac{(-1)^j q\left(\begin{array}{c} q-1 \\ j-2 \\ \end{array} \right)}{j(j-1)} (q(j^2-3j+1)+1) \;, \nonumber \\
K_j^{q+1,q}(q) &=& \frac{(-1)^j q\left(\begin{array}{c} q-1 \\ j-2 \\ \end{array} \right)}{j(j-1)} (1-q(j-1))\;, \nonumber \\
K_j^{q+1,q}(q+1) &=& \frac{(-1)^j q\left(\begin{array}{c} q-1 \\ j-2 \\ \end{array} \right)}{j(j-1)} (q+1)\;. \nonumber
\end{eqnarray}

\noindent
Thus, a direct substitution of the previous four expressions in (\ref{eqdual1}) proves (\ref{eqdual2}). Finally, note that 

$$2(q-1)^{j-1}>(-1)^j((j-1)q((j-2)q-2)+2)\;,$$ 

\noindent
for all $q\geq 5$, and $4\leq j \leq q+1$. Therefore, $A_1^{\perp}=A_2^{\perp}=A_3^{\perp}=0$ and $A_j^{\perp}>0$, for all $4\leq j \leq q+1$. Consequently, ${\cal C}_{(1,q+1,q^2)}^{\perp}$ is a $(q-2)$-weight cyclic code if $q\geq 5$. 
\end{proof}

\begin{remark}
By using (\ref{eqdual2}) it is easy to obtain

$$A_5^{\perp}=\frac{(q^2-1)q(q-1)(q-2)(q-3)(q-4)}{120}\;.$$

\noindent
Thus, note that $A_5^{\perp}=0$, when $q=4$. In fact, in this case, since dimension of ${\cal C}_{(1,5,4^2)}^{\perp}$ is $q-2=2$ and $u=\gcd(q+1,\frac{q^2-1}{n})=\gcd(q+1,q-1)=1$, Theorem \ref{teoCCI} tell us that ${\cal C}_{(1,5,4^2)}^{\perp}$ is a one-weight irreducible cyclic code of length 5, whose nonzero weight is $4$. That is, ${\cal C}_{(1,q+1,q^2)}^{\perp}$ is not a $(q-2)$-weight cyclic code, when $q=4$.
\end{remark}

The following are some examples of Theorems \ref{mainR1} and \ref{mainR2}.

\begin{example}\label{ejetres}
Let $q=5$. Thus, by Theorems \ref{mainR1} and \ref{mainR2}, we can see that both ${\cal C}_{(1,6,5^2)}$ and ${\cal C}_{(1,6,5^2)}^{\perp}$ are optimal three-weight cyclic codes of length $6$ and dimension $3$, whose weight enumerator polynomial is

\[ 1+60z^{4}+24z^{5}+40z^{6} \; . \]
\end{example}

\begin{example}\label{ejecuatro}
Let $q=7$. Thus, by Theorem \ref{mainR1}, we can see that ${\cal C}_{(1,8,7^2)}$ is an optimal three-weight cyclic code of length $8$ and dimension $3$, whose weight enumerator polynomial is

\[ 1+168z^{6}+48z^{7}+126z^{8} \; . \]

\noindent
On the other hand, by Theorem \ref{mainR2}, ${\cal C}_{(1,8,7^2)}^{\perp}$ is an optimal five-weight cyclic code of length $8$ and dimension $5$, where $A_4^{\perp}=420$. In fact, by means of (\ref{eqdual2}), the weight enumerator polynomial of ${\cal C}_{(1,8,7^2)}^{\perp}$ is

\[ 1+420z^{4}+1008z^{5}+4032z^{6}+6432z^{7}+4914z^{8} \; . \]

\end{example}

\begin{example}\label{ejecinco}
Let $q=8$. Thus, by Theorem \ref{mainR1}, we can see that ${\cal C}_{(1,9,8^2)}$ is an optimal three-weight cyclic code of length $9$ and dimension $3$, whose weight enumerator polynomial is

\[ 1+252z^{7}+63z^{8}+196z^{9} \; . \]

\noindent
On the other hand, by Theorem \ref{mainR2}, ${\cal C}_{(1,9,8^2)}^{\perp}$ is an optimal six-weight cyclic code of length $9$ and dimension $6$, where $A_4^{\perp}=882$. In fact, by means of (\ref{eqdual2}), the weight enumerator polynomial of ${\cal C}_{(1,9,8^2)}^{\perp}$ is

\[ 1+882z^{4}+3528z^{5}+19992z^{6}+57456z^{7}+101493z^{8}+78792z^{9} \; . \]

\end{example}

\begin{example}\label{ejeseis}
Let $q=9$. Thus, by Theorem \ref{mainR1}, we can see that ${\cal C}_{(1,10,9^2)}$ is an optimal three-weight cyclic code of length $10$ and dimension $3$, whose weight enumerator polynomial is

\[ 1+360z^{8}+80z^{9}+288z^{10} \; . \]

\noindent
On the other hand, by Theorem \ref{mainR2}, ${\cal C}_{(1,10,9^2)}^{\perp}$ is an optimal seven-weight cyclic code of length $10$ and dimension $7$, where $A_4^{\perp}=1680$. In fact, by means of (\ref{eqdual2}), the weight enumerator polynomial of ${\cal C}_{(1,10,9^2)}^{\perp}$ is

\[ 1+1680z^{4}+10080z^{5}+77280z^{6}+343680z^{7}+1036440z^{8}+1840880z^{9}+1472928z^{10} \; . \]

\end{example}

\begin{remark}
By directly obtaining the corresponding weight distribution, the previous numerical examples were corroborated with the help of a computer.
\end{remark}

\section{Conclusions}\label{conclusiones}
By describing the number of occurrences of the symbols in the codewords of a particular kind of one-weight and semiprimitive two-weight irreducible cyclic codes of dimension 2 (Proposition \ref{prdos}), we present a new class of optimal three-weight cyclic codes of dimension 3 (Theorem \ref{mainR1}). Furthermore, we explicitly determine the weight distribution for the cyclic codes in this new class and, with the knowledge of this distribution, we study their corresponding dual codes. As a result, we found that, except for the case $q=2$, all the dual codes have minimum Hamming distance $4$. In consequence, since the length, dimension and minimum Hamming distance of the dual codes are $q+1$, $q-2$, and $4$, respectively, our last result (Theorem \ref{mainR2}) shows that these codes are also optimal. As an application of the Krawtchouck polynomials we obtained the weight distribution of the dual codes, making it clear that all of them are $(q-2)$-weight cyclic codes for $q\geq 5$. Therefore, in addition to our new class of optimal cyclic codes, this last result gives access to an infinite family of optimal $(q-2)$-weight cyclic codes of length $q+1$, and dimension $q-2$, whose minimum Hamming distance is $4$. In fact, note that the redundancy for each one of the dual codes is only $3$. However, despite this low redundancy, any dual code is able to correct an error and detect up to two errors.

Finally, we want to note that the dual code in Example \ref{ejetres} is an optimal three-weight cyclic code of dimension 3 that is not included in the class of optimal three-weight cyclic code studied here (the dual code is not of the form ${\cal C}_{(1,q+1,q^2)}$). Therefore, this example shows that the results presented in this work are likely to be generalized. 

\bibliographystyle{IEEE}

\begin{thebibliography}{1}
\bibitem{Anderson} R. Anderson, C. Ding, T. Helleseth, and T. Kl$\o$ve, ``How to build robust shared control systems," {\it Designs, Codes Cryptogr.}, vol. 15, no. 2, pp. 111-124, 1998.

\bibitem{Calderbank} A. R. Calderbank and J. M. Goethals, ``Three-weight codes and association schemes," {\it Philips J. Res.}, vol. 39, nos. 4-5, pp. 143-152, 1984.

\bibitem{Delsarte} P. Delsarte, ``On subfield subcodes of Reed-Solomon codes," {\it IEEE Trans. Inf. Theory}, vol. 21, no. 5, pp. 575-576, 1975.

\bibitem{Griesmer} J. H. Griesmer, ``A bound for error correcting codes," {\it IBM Journal of Res. and Dev.}, vol. 4, no. 5, pp. 532-542, 1960.

\bibitem{Heng} Z. Heng and Q. Yue, ``Several Classes of Cyclic Codes With Either Optimal Three Weights or a Few Weights," {\it IEEE Trans. Inf. Theory}, vol. 62, no. 8, pp. 4501-4513, 2016.

\bibitem{Huffman} W. C. Huffman and V. Pless, {\em Fundamentals of Error-Correcting Codes}. Cambridge, U.K.: Cambridge Univ. Press, 2003.

\bibitem{MacWilliams}  F. J. MacWilliams and N. J. A. Sloane, {\em The Theory of Error-Correcting Codes}. Amsterdam, The Netherlands: North-Holland, 1977.

\bibitem{Pless} V. Pless, ``Power moment identities on weight distributions in error-correcting codes," {\it Inf. Contr.}, vol. 6, pp. 147-152, 1962.

\bibitem{Schmidt} B. Schmidt and C. White, ``All two-weight irreducible cyclic codes?," {\it Finite Fields Appl.}, vol. 8, pp. 1-17, 2002.

\bibitem{Solomon} G. Solomon and J. J. Stiffler, ``Algebraically punctured cyclic codes," {\it Inform. and Control}, vol. 8, no. 2, pp. 170-179, 1965.

\bibitem{Vega1} G. Vega, ``A characterization of a class of optimal three-weight cyclic codes of dimension 3 over any finite field," {\it Finite Fields Appl.}, vol. 42, pp. 23-38, 2016.

\bibitem{Vega2} G. Vega, ``A characterization of all semiprimitive irreducible cyclic codes in terms of their lengths," {\it Applicable Algebra Eng. Commun. Computing}, vol. 30, no. 5, pp. 441-452, 2019.

\bibitem{Wolfmann} J. Wolfmann, ``Are 2-Weight Projective Cyclic Codes Irreducible?," {\it IEEE Trans. Inform. Theory}, vol. 51, no. 2, pp. 733-737, 2005.
\end{thebibliography}

\end{document}